\documentclass[12pt,english,floatfix,showkeys,superscriptaddress,aps,prd,preprint]{revtex4}
\usepackage[latin1]{inputenc}
\usepackage[T1]{fontenc}
\usepackage{lmodern}
\usepackage{amsmath}
\usepackage{amssymb}
\usepackage{graphicx}
\usepackage{caption}
\usepackage{subcaption}
\usepackage{float}
\usepackage{esint}
\usepackage{dcolumn}
\usepackage{babel}
\usepackage{csquotes}
\usepackage{color}
\usepackage{slashed}
\usepackage{simplewick}
\usepackage{amsmath,latexsym}

\usepackage{hyperref}
\hypersetup{
    colorlinks,
    citecolor=blue,
    filecolor=green,
    linkcolor=purple,
    urlcolor=red,
}

\usepackage{slashed}

\usepackage{hyperref}
\hypersetup{colorlinks,breaklinks,
			citecolor=[rgb]{0,0.0,1.0},
            urlcolor=[rgb]{0.0,0.0,1.0},
            linkcolor=[rgb]{0,0.5,0.9}}

\begin{document}

\title{Traversable Wormholes from Loop Quantum Gravity}

\author{M. B. Cruz}
\email{messiasdebritocruz@servidor.uepb.edu.br}
\affiliation{Universidade Estadual da Para\'iba (UEPB), \\ Centro de Ci\^encias Exatas e Sociais Aplicadas (CCEA), \\ R. Alfredo Lustosa Cabral, s/n, Salgadinho, Patos - PB, 58706-550 - Brazil.}
\author{R. M. P. Neves}
\email{raissa.pimentel@uece.br}
\affiliation{Universidade Estadual do Cear\'a (UECE), Faculdade de Educa\c{c}\~ao, Ci\^encias e Letras de Iguatu, Av. D\'ario Rabelo s/n, Iguatu - CE, 63.500-00 - Brazil.}
\author{Celio R. Muniz}
\email{celio.muniz@uece.br}
\affiliation{Universidade Estadual do Cear\'a (UECE), Faculdade de Educa\c{c}\~ao, Ci\^encias e Letras de Iguatu, Av. D\'ario Rabelo s/n, Iguatu - CE, 63.500-00 - Brazil.}



\date{\today}

\begin{abstract}
This study introduces and investigates Lorentzian traversable wormhole solutions rooted in Loop Quantum Gravity (LQG). The static and spherically symmetric solutions to be examined stem from the energy density sourcing self-dual regular black holes discovered by L. Modesto, relying on the parameters associated with LQG, which account for the quantum nature of spacetime. We specifically focus on macroscopic wormholes characterized by small values of these parameters. Our analysis encompasses zero-tidal solutions and those with non-constant redshift functions, exploring immersion diagrams, curvatures, energy conditions, equilibrium requirements, and the requisite quantity of exotic matter to sustain these wormholes. The investigation underscores the influence of LQG parameters on these features, highlighting the pivotal role of spacetime's quantum properties in shaping these wormholes and governing their behavior.
\end{abstract}


\maketitle

\section{Introduction}

General Relativity (GR) is widely acknowledged as the most accurate description of gravitational physics. Among the most intriguing predictions of Einstein's theory are the propositions of black holes (BHs) \cite{Oppenheimer:1939ue} and wormholes (WHs) \cite{Hawking:1988ae, Morris:1988cz}. Wormholes are hypothetical objects that function as space-time bridges, connecting two distinct points in space-time across the universe, potentially serving as shortcuts for space-time travel between them \cite{Einstein:1935tc, Morris:1988tu, Frolov:2023res}. Despite being one of the most captivating predictions of GR, wormholes have not yet been directly detected through observations. In contrast, black holes have been directly observed through gravitational waves (GWs) observations, originating from the merging of binary BHs. This class of events has been observed with remarkable precision by the LIGO and Virgo collaborations \cite{LIGOScientific:2016aoc}.

These objects represent potential phenomena that could aid in unraveling one of the most perplexing issues in contemporary theoretical physics: the nature of quantum gravity. This is because, in the presence of an intensely strong gravitational field, the quantum properties of spacetime must come into play. Currently, two primary contenders for a theory of quantum gravity (QG) are Loop Quantum Gravity (LQG) \cite{Rovelli:2014ssa, Rovelli:2003wd}, in competition with String Theory (ST) \cite{Zwiebach:2004tj}. Within the framework of LQG, it is possible to develop intriguing theoretical models shedding light on the quantum characteristics of spacetimes, as revealed by BHs \cite{Modesto:2005zm, Peltola:2008pa, Gambini:2013ooa, Olmedo:2016ddn} and WHs in Loop Quantum Cosmology (LQC) \cite{Sengupta:2023yof}. An important scenario pertains to the quantum version of the Schwarzschild black hole (SchBH), known as a self-dual black hole (SDBH) \cite{Modesto:2009ve}. Significant studies have been conducted on the SDBH solution, presenting compelling properties \cite{Brown:2010csa, Silva:2017gki, Cruz:2015bcj, Cruz:2020emz, Santos:2021wsw}. For example, the physical singularity of an SDBH can be replaced by an asymptotically flat region, an anticipated effect in the realm of quantum gravity.

As mentioned earlier, WHs have not been directly detected. However, delving into the fascinating aspects of these objects can enrich our comprehension of GR and, consequently, contribute to addressing one of the most perplexing problems: the quest for a viable theory of QG. In this context, WHs that are asymptotically flat at infinitely large radial distances from the throat do exist as static, spherically symmetric solutions to the Einstein field equations \cite{Fuller:1962zza}. Nevertheless, realistic Schwarzschild WHs prove to be unstable at the throat due to the development of infinitely large gravitational tidal forces, resulting in a Weyl curvature singularity at the throat. This generation of substantial tidal forces is attributed to the gravitational attraction of matter at the throat by the two extremities in opposite directions of the wormholes. Morris and Thorne proposed a solution to this problem in Refs. \cite{Morris:1988cz, Morris:1988tu}, suggesting that the matter at the throat be replaced by a form of exotic gravitationally repulsive matter. However, it's essential to note that some considerations arise, as the energy density of such matter always remains positive. Therefore, the Null Energy Condition (NEC), expressed as $\rho + p_i \geq 0$, must be violated by the content of matter at the throat.

However, it is crucial to note that, while a violation of the Strong Energy Condition (SEC), expressed as $\rho + p_r+2 p_t \geq 0$, is necessary to achieve an accelerating universe \cite{SupernovaSearchTeam:1998fmf, Dalal:2000xw}, the violation of the NEC takes precedence. For example, such a violation may lead to quantum instabilities in the vacuum \cite{Cai:2009zp}. Another critical aspect concerning WHs is the radial metric potential of the static, spherically symmetric metric, known as the shape function $b(r)$. According to the prescription in Refs. \cite{Morris:1988cz, Morris:1988tu}, the shape function determines the geometry of the WHs and must adhere to the following criteria: (i) the $b(r)$ at the throat radius, $r_0$, must be equal to the throat radius itself, i.e., $b(r_0) = r_0$. (ii) For radial distances $r > r_0$, the ratio of the shape function to the radial distance must be less than unity, expressed as $b(r)/r < 1$. (iii) The derivative of the shape function concerning the radial distance at the throat must be less than unity, i.e., $b'(r)_{r=r_0} < 1$. Finally, (iv) implies a minimum size for the throat, which, in turn, minimizes the amount of exotic matter required at the throat to violate the NEC.

To violate any of the energy conditions, one can achieve this by modifying the sectors of matter or geometry in the Einstein field equations \cite{Zlatev:1998tr, Urena-Lopez:2002nup, Dvali:2000hr, Nojiri:2006ri}. Traversable WHs have been constructed in the literature using both approaches \cite{Das:2022wzp, Estrada:2023pny, Santos:2023zrj}. The hypothesis of the existence of these objects in certain regions of the galaxies, including ours, has been recently explored in \cite{Barcelo:1999hq, Hayward:2002pm, Sengupta:2023ysx, Eiroa:2005pc, Rosa:2022osy, Muniz:2022eex}. The LQG-inspired effective models are recognized for addressing the strong curvature singularity at the center of BHs and the Big Bang singularity. In both cases, the singularity problem was found to be resolved \cite{Modesto:2009ve, Ashtekar:2008ay}. Consequently, this serves as a strong motivation for attempting to construct WHs using these models, where the curvature singularity may be resolved at the wormhole throat, enabling their traversability.

This work aims to build novel traversable, static, and spherically symmetric WH solutions based on the effective energy density that sources self-dual regular black holes studied in the framework of LQG theory. Although the horizons of these latter are hidden behind a wormhole of Planck diameter \cite{Modesto:2009ve}, we intend to show here the feasibility of the existence of independent macroscopic wormholes in that context. The objective is to enhance the understanding of the relationship between quantum gravity effects and the emergence of exotic spacetime structures. The paper is organized as follows: In Section II, we present a brief description of the SDBH and the effective energy density that sources it, which will serve as a basis for finding the proposed WH solutions.  Within the same section, we include subsections detailing different scenarios where we conduct analysis, considering different values of LQG parameters, related to the corresponding immersion diagrams, curvatures, energy conditions, and the required quantity of exotic matter to sustain them, as well as equilibrium conditions. We conclude by summarizing our results and offering concluding remarks in Section III. Throughout this paper, we utilize natural units with $8 \pi G = 1$ and adopt the metric signature $(-, +, +, +).$


\section{LQG-Inspired Wormholes}

In this section, our exploration of traversable wormholes is inspired by principles from LQG. By doing so, we aim to uncover intriguing insights arising from the integration of quantum gravity concepts with the existence of WHs. Thus, we first provide a brief introduction to the essential aspects emerging from a simplified model of LQG. This model involves an asymmetry-reduced approach corresponding to homogeneous spacetimes within a black hole context \cite{Modesto:2009ve}.

In this context, we interpret the SDBH solution as arising from an effective matter fluid that simulates corrections within the framework of LQG \cite{Modesto:2009ve, Bonanno:2000ep}. Consequently, the effective gravity-matter system obeys the Einstein equation:
\begin{eqnarray}
    G^{\mu}_{\ \nu} \equiv R^{\mu}_{\ \nu} - \frac{1}{2} g^{\mu}_{\ \nu} R = T^{\mu}_{\ \nu} ,
\end{eqnarray}
where $T_{\mu \nu}$ denotes the effective energy-momentum tensor. The energy-momentum tensor corresponding to a perfect fluid that is compatible with the SDBH is:
\begin{eqnarray}
    T^{\mu}_{\ \nu} = \text{diag}\left(-\rho, p_r, p_t, p_t \right),
\label{energy_tensor}
\end{eqnarray}
where $\rho = - G^{t}_{\ t}$, $p_r = G^{r}_{\ r}$ and $p_{t} = G^{\theta}_{\theta} = G^{\phi}_{\phi}$. The analytical expression for the energy density is given by \cite{Modesto:2009ve}:
\begin{eqnarray}
\rho(r) &&= 4 r^4 \Bigg[a_0^4 \frac{r_0}{2} (1 + {\mathcal P})^2 + \frac{r_0^2}{2} {\mathcal P} (1 + {\mathcal P})^2  r^7  - a_0^2  r^2 (r_0 {\mathcal P} + r) (3 r_0^2 {\mathcal P}^2 \nonumber \\ && \hspace{0.3cm} - \frac{r_0}{2} (7 + {\mathcal P} (2 + 7 {\mathcal P})) r + 3 r^2) \Bigg] \Bigg{/} \Bigg[ (r_0 {\mathcal P} + r)^3 (a_0^2 + r^4)^3\Bigg],
\label{edensity}
\end{eqnarray}
here, it is important to note that we have replaced the SDBH mass with the WH throat radius $r_0$, i.e., $m=r_0/2$. In Eq. \eqref{edensity}, the LQG parameters appear, and the polymeric function $\mathcal{P}$ is given by:
\begin{eqnarray}
    \mathcal{P} = \frac{\sqrt{1+\epsilon^2} - 1}{\sqrt{1+\epsilon^2} + 1} .
\label{P_parameter}
\end{eqnarray}
In Eq. \eqref{P_parameter}, the parameter $\epsilon$ is defined as $\epsilon = \gamma \delta_b$, where $\gamma$ is the Barbero-Immirzi parameter, and $\delta_b$ is the polymeric parameter used for quantization in LQG. Still, in Eq. \eqref{edensity}, the parameter $a_0$ is introduced, defined by $a_0 \equiv A_{\text{min}}/8\pi$, where $A_{\text{min}}$ represents the minimal area in the context of LQG.

Therefore, moving forward, we will investigate the wormholes in the context of Loop Quantum Gravity (LQG). We will particularly focus on the energy density given in Eq. \eqref{edensity} by the static and spherically symmetric Morris-Thorne wormhole metric as presented by \cite{Morris:1988cz}:
\begin{equation}\label{metric1}
    ds^2=-e^{2\Phi(r)}dt^2+\frac{dr^2}{1-\frac{b(r)}{r}}+r^2d\Omega_2,
\end{equation}
Here, $\Phi(r)$ represents the redshift function, $b(r)$ is the shape function, and $d\Omega_2=d\theta^2+\sin^2\theta d\phi^2$ denotes the spherical line element. Given the metric ansatz of Eq. \eqref{metric1}, Einstein's equations take on their simplest form:
\begin{align}
G_{\ t}^{t}= & \frac{b'}{r^{2}}=\rho(r),\label{eq:g00}\\
G_{\ r}^{r}= & -\frac{b}{r^{3}}+ 2\frac{\left(r-b\right)\Phi'}{r^{2}}= p_{r}(r),\label{eq:grr}\\
G_{\theta}^{\theta}=G_{\phi}^{\phi}= & \left(1-\frac{b}{r}\right)\left[\Phi''+(\Phi')^{2}+\frac{\left(b-rb'\right)}{2r(r-b)}\Phi'+\frac{\left(b-rb'\right)}{2r^2(r-b)}+\frac{\Phi'}{r}\right]= p_{t}(r),\label{eq:gthetatheta}
\end{align}
where we utilize the effective energy-momentum tensor from Eq. \eqref{energy_tensor}, where $\rho(r)$ represents the surface energy density, $p_{r}(r)$ is the radial pressure, and $p_{t}(r)$ denotes the lateral pressures. Hence, the quantity $\rho(r)$ described by Eq. \eqref{edensity} will be regarded as the energy source for the new wormholes investigated in various scenarios in the following subsections.

\subsection{Case $\Phi'(r)=0$}

Initially, we will explore the simplest scenario with $\Phi(r)=$ constant (a zero-tidal traversable wormhole). By solving the time component of Einstein's equations with the metric provided in Eq. (\ref{metric1}) and the energy density given by Eq. (\ref{edensity}), we can derive a closed shape function. However, due to its complexity, we will resort to an approximation suitable for small LQG parameters. 

This full solution prompts a critical evaluation of the necessary conditions that must be met for a traversable wormhole, specifically: $(i)$  asymptotic flatness, {\it i.e.}, $b(r)/r\to 0$ for $r\to \infty$; $(ii)$ the flaring-out condition, determined by the minimality of the wormhole throat, such that $(b-b'r)/b^{2}>0 \Rightarrow b'(r_0)<1$, and at the throat, $b(r_{0})=r_{0}$; $(iii)$ the condition $1-b/r\geq0$ must also satisfy to guarantee wormhole solutions; $(iv)$ it must be ensured that there are no horizons present, which are identified as surfaces with $e^{\Phi}\rightarrow0$, so that $\Phi(r)$ is finite everywhere \cite{Morris:1988cz}.  

\begin{figure}[!h]
 \centering
    \includegraphics[width=0.58\textwidth]{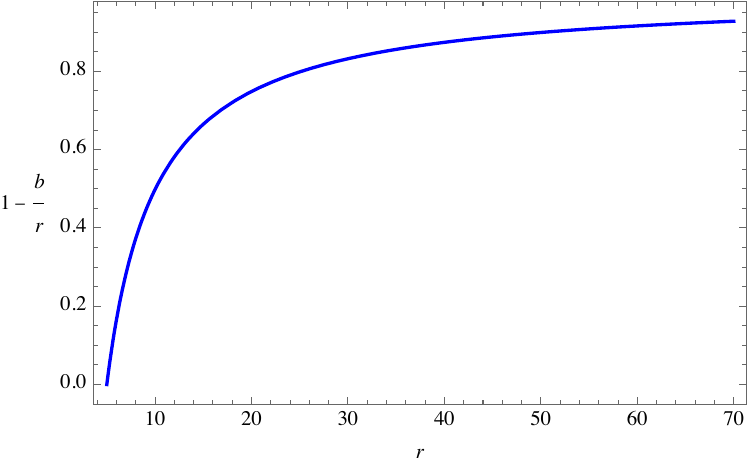}
        \caption{The function $1-b(r)/r$, for $a_0=1.0, \mathcal{P}=0.1$ and $r_0=5.0$.}
    \label{A}
\end{figure}

We can demonstrate that the obtained solution satisfies condition $(i)$ regardless of the values assigned to the parameters. The flaring-out condition $(ii)$, is met based on the inequality 
\begin{equation}
    b'(r_0)=\frac{2 r_0^4(a_0^2+\mathcal{P} r_0^4)}{(1+\mathcal{P})(a_0^2+r_0^4)^2}<1,
\end{equation}
under the trivial conditions that $\mathcal{P}$, $r_0$, and $a_0$ are all greater than zero. Moreover, the fulfillment of condition $(iii)$ is affirmed based on the findings depicted in Figure \ref{A}, which accounts the condition $(i)$, too. Similar graphical representations are consistently obtained across various parameter values.
Finally, since we are considering zero-tidal wormholes, the condition $(iv)$ is automatically satisfied. From the comprehensive solution found for the shape function, a quite reasonable approximation can be derived by considering significantly small LQG parameters, {\it i.e.} when $\sqrt{a_0}\ll r_0$ and $\mathcal{P}\ll 1$. This assumption finds support in astrophysical observations \cite{Yang:2023gas}. Thus, we find a simpler shape function given by 
\begin{equation}\label{SolApp}
    b(r)\approx  r_0\left(1+2\mathcal{P}\right)+\frac{a_0^2}{r_0^3}\left(\frac{\mathcal{P}}{5}-\frac{1}{2}\right)-\frac{2\mathcal{P}r_0^2}{r}+\frac{4 a_0^2}{ r^3}-\frac{7 a_0^2 r_0}{ r^4}\left(\mathcal{P}+\frac{1}{2  }\right)+\frac{34 a_0^2 \mathcal{P} r_0^2}{5  r^5},
\end{equation}
upon expanding up to $\mathcal{O}(a_0^2)$ and $\mathcal{O}(\mathcal{P})$. This macroscopic wormhole solution is asymptotically flat and satisfies $b(r_0)=r_0$. Additionally, it adheres to flaring-out conditions ($b'(r_0)<1$) when $0<\mathcal{P}<(2 a_0^2-r_0^4)/(6 a_0^2-2 r_0^4)$ and $a_0<r_0^2\sqrt{3}/3$, which are easily satisfied given the small values of the LQG parameters, so that the upper bound for $\mathcal{P}$ is $1/2$, upon admitting $r_0\gg \sqrt{a_0}$. Condition ($iv$) is also met across a broad spectrum of LQG parameters and $r_0$ radii. Henceforth, we will accept the shape function provided by Eq. (\ref{SolApp}) as a given and proceed to examine the wormhole solutions derived from it. 

\begin{figure}[!h]
\centering
\begin{subfigure}{.5\textwidth}
  \centering
  \includegraphics[width=0.85\linewidth]{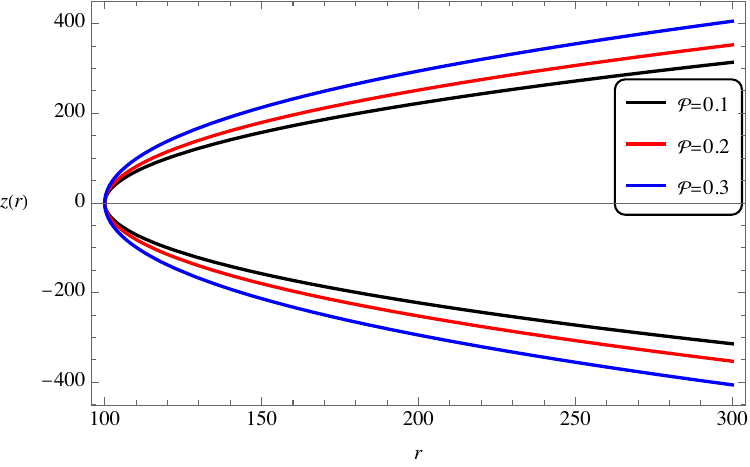}
\end{subfigure}%
\begin{subfigure}{.5\textwidth}
  \centering
  \includegraphics[width=1.05\linewidth]{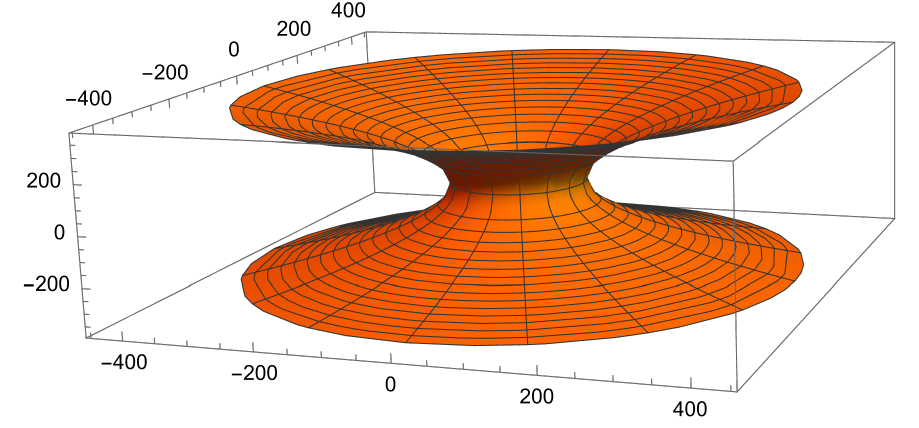}
\end{subfigure}
\caption{Left Panel: Profiles of the embedding diagrams related to LQG-inspired wormholes, as functions of the radial coordinate, for $r_0=100$, $a_0=1.0$, and some values of $\mathcal{P}$. Right panel: 3-D embedding diagram, considering $\mathcal{P}=0.1$ and the same remaining parameters.}
\label{Emb}
\end{figure}

\begin{figure}[H]
 \centering
   \includegraphics[width=0.6\textwidth]{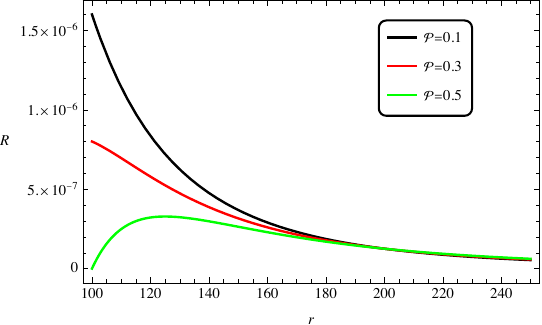}
      \caption{Ricci's curvature scalar as a function of the radial coordinate, for some values of $\mathcal{P}$, with $a_0=1.0$, $r_0=100$.}
  \label{RA}
\end{figure}

In Figure \ref{Emb}, we plot some embedding diagrams for the LQG-inspired wormholes. The left panel illustrates a noteworthy trend, namely, as the LQG parameter $\mathcal{P}$ increases, noticeable deviations from flatness occur in the wormhole mouth, which corresponds to growing slopes toward the throat. In other words, the augmenting of $\mathcal{P}$ leads to pronounced quantum effects, causing accentuated distortions in the geometry of the wormhole.
Let us see if this statement is corroborated by the analysis of Ricci's curvature scalar. For the obtained zero-tidal wormhole solution this quantity is depicted in Figure \ref{RA} as a function of the radial coordinate for some values of $\mathcal{P}$. We can see that, as this LQG parameter increases, the wormhole solution shows reduced curvatures. However, it tends to display a locally accentuated maximum near the throat. These features differ from the subsequent cases we will investigate, where the redshift function is non-constant and contributes to the curvature.

Now we analyze the energy conditions that the physical systems should preferentially satisfy, namely: Null Energy Conditions (NEC: $\rho+p_i\geq 0$, with $i=\{r,t\}$), Weak Energy Conditions (WEC: NEC plus $\rho\geq 0$) as well as Strong Energy Conditions (SEC: WEC plus $\rho+p_r+2p_t\geq 0$). For the solution given by Eq. \eqref{SolApp} we have, according to Einstein's equations the  quantities
\begin{eqnarray}
   \rho&=& \frac{2}{  r^8}(4 a_0^2 \mathcal{P}  r_0 r-17 a_0^2 \mathcal{P} r_0^2-6 a_0^2 r^2+7 a_0^2  r_0 r+14 a_0^2\mathcal{P}r_0 r+\mathcal{P}  r_0^2 r^4),\label{rho}\\
   p_r&=& -\frac{1}{10 r_0^3 r^8}(-5 a_0^2 r^5 + 2 a_0^2 \mathcal{P} r^5 + 
   40 a_0^2  r_0^3 r^2- 35 a_0^2 r_0^4  r - 70 a_0^2 \mathcal{P} r_0^4  r\nonumber\\
   &+&   20 \mathcal{P} r_0^4  r^5 + 10 r_0^4  r^5  + 68 a_0^2 \mathcal{P} r_0^5 - 20 \mathcal{P}  r_0^5 r^4),\label{p_r}\\
   p_t&=&\frac{1}{20  r_0^3 r^8}(-5 a_0^2 r^5 + 2 a_0^2 \mathcal{P} r^5 + 160 a_0^2 r_0^3  r^2 - 175 a_0^2  r_0^4 r- 
  350 a0^2 \mathcal{P}  r_0^4 r\nonumber\\
  &+& 20 \mathcal{P}  r_0^4 r^5 + 10  r_0^4 r^5 + 
  408 a_0^2 \mathcal{P} r_0^5 - 40 \mathcal{P}  r_0^5 r^4).
  \label{p_t}
\end{eqnarray}

In Figure \ref{Fig2}, the density of the source is depicted alongside its correlations with pressures. The breach of the energy conditions, specifically due to the violation of the Null Energy Condition (NEC), is observed. This breach is alleviated by an escalation in the LQG parameter $\mathcal{P}$, particularly noticeable near the throat as illustrated in the right panel. This observation suggests that as space diverges from classical structures and gradually adopts quantum properties, it tends to more closely adhere to these conditions.
\begin{figure}[!h]
\centering
\begin{subfigure}{.5\textwidth}
  \centering
  \includegraphics[width=1.01\linewidth]{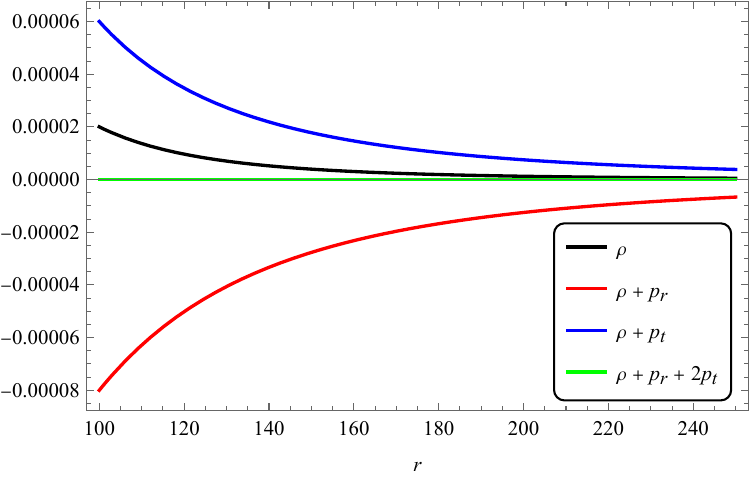}
\end{subfigure}%
\begin{subfigure}{.5\textwidth}
  \centering
  \includegraphics[width=1.01\linewidth]{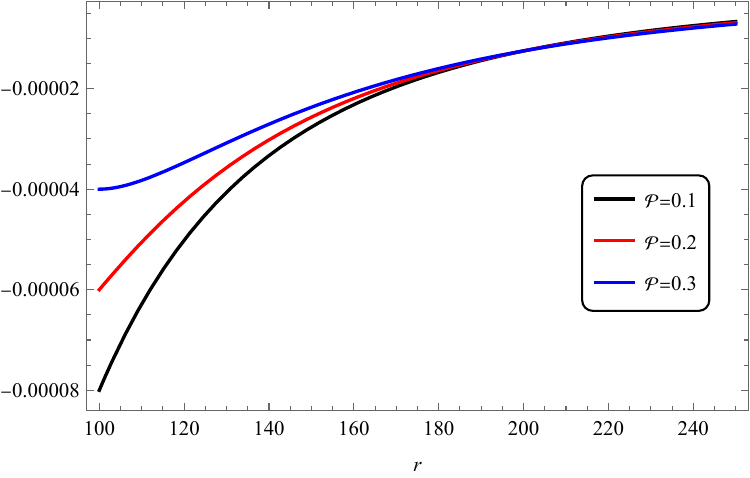}
\end{subfigure}
\caption{Left Panel: Energy density and its combinations with the pressures, as functions of the radial coordinate, for $r_0=100$, $a_0=1.0$, and $\mathcal{P}=0.1$. Right panel: Energy density plus radial pressure, in detail, for the same parameters and other values of $\mathcal{P}$.}
\label{Fig2}
\end{figure}

We are now going to analyze the Volume Integral Quantifier (VIQ), defined by \cite{Nandi:2004}
\begin{equation}\label{viq}
\mathcal{I}_q=2\int_{r_0}^x 4\pi r^2 (\rho+p_r)dr,
\end{equation}
where $x$ is very close to $r_0$. The objective is to obtain the quantity of exotic matter necessary to keep the wormhole throat open. Plugging Eqs. (\ref{rho}) and (\ref{p_r}) into (\ref{viq}), and considering $x=r_0+\sqrt{a_0}$, we obtain a quite involved expression. However, the referred quantity converges to a simpler expression for large wormholes; in other words, $\mathcal{I}_q\to -8\pi\sqrt{a_0}(1-2\mathcal{P})$ as $r_0$ approaches infinity. Observe that the requirement for exotic matter diminishes as $\mathcal{P}$ increases, {\it i.e.}, when the quantum behavior of spacetime becomes more robust, with VIQ tending to vanish in the upper limit $\mathcal{P}\to 1/2$.

Another feature that we can examine is the wormhole stability via Tolman-Oppenheimer-Volkoff (TOV) equation, given by
\begin{equation}\label{TOV}
    -\frac{dp_r}{dr}-\Phi'(\rho+p_r)+\frac{2}{r}(p_t-p_r)=0,
\end{equation}
where the left-hand two first terms are identified with the hydrostatic ($F_h$) and gravitational ($F_g$) forces, respectively, and the last one is associated with the anisotropic ($F_a$) force. As we are considering the zero-tidal wormhole in which $\Phi'=0$, substituting Eqs. (\ref{rho}), (\ref{p_r}), and (\ref{p_t}) in (\ref{TOV}), we find that the equilibrium condition is obeyed, namely, $F_h+F_a=0$.

\subsection{Case $\Phi(r) = \frac{r_0}{r}$}

In this case, in which the redshift function does not depend on the LQG parameters in a trivial manner, we get the radial and transversal pressures:
\begin{eqnarray}
    p_r &=& \frac{r_0}{r^5} \Bigg[2 r (3 \mathcal{P} r_0+r_0)-\big((2 \mathcal{P}+3) r^2\big)-4 \mathcal{P} r_0^2\Bigg]-\frac{1}{10 r^9 r_0^3} \Bigg[a_0^2 (r-2 r_0) (r-r_0)  \nonumber \\ &\times&\big((2 \mathcal{P}-5) r^4+(2 \mathcal{P}-5) r^3 r_0+(2 \mathcal{P}-5) r^2 r_0^2+(2 \mathcal{P}+35) r r_0^3-68 \mathcal{P} r_0^4\big)\Bigg] ,
\end{eqnarray}
\begin{eqnarray}
    p_t &=& \frac{r-r_0}{20 r^{10} r_0^3} \Bigg[a_0^2 \big((2 \mathcal{P}-5) r^6+2 (5-2 \mathcal{P}) r^5 r_0+4 (5-2 \mathcal{P}) r^4 r_0^2+4 (45-2 \mathcal{P}) r^3 r_0^3 \nonumber \\ &-& (358 \mathcal{P}+235) r^2 r_0^4+10 (54 \mathcal{P}-7) r r_0^5+136 \mathcal{P} r_0^6\big)+10 r^4 r_0^4 \big((2 \mathcal{P}+3) r^2 \nonumber \\ &+& (2-8 \mathcal{P}) r r_0-4 \mathcal{P} r_0^2\big) \Bigg] ,
\end{eqnarray}
and the energy density, $\rho(r)$, is the same as the Eq. \eqref{p_t}.
\begin{figure}[!h]
\centering
\begin{subfigure}{.5\textwidth}
  \centering
  \includegraphics[width=0.9\linewidth]{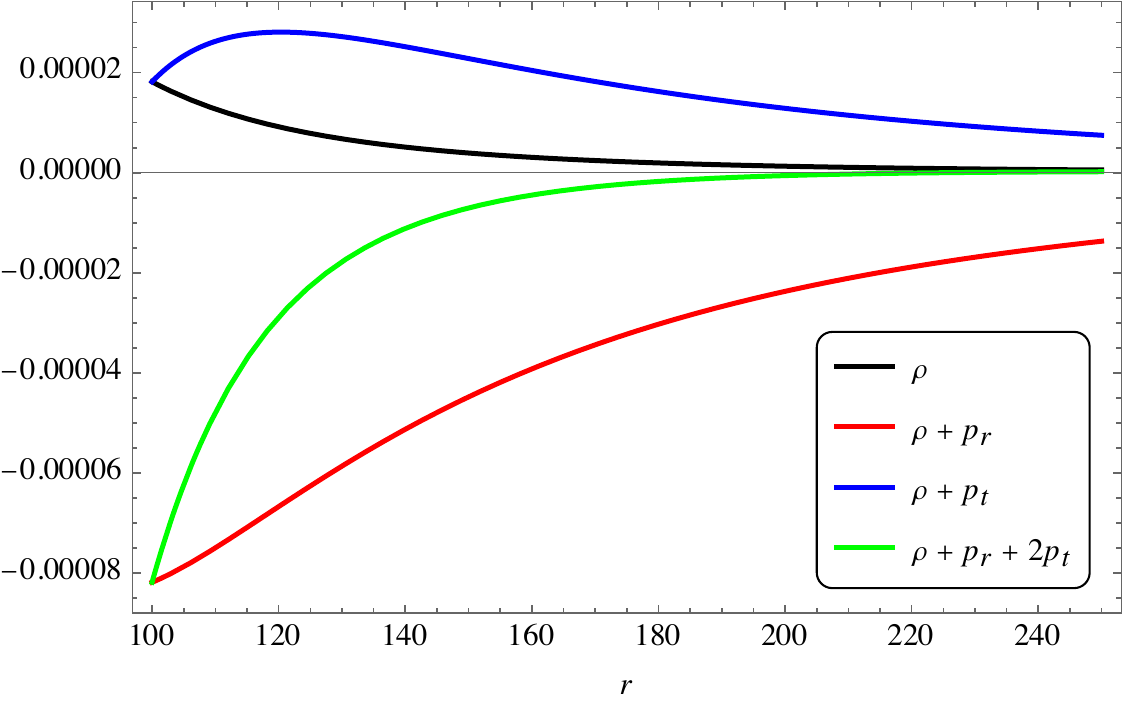}
\end{subfigure}%
\begin{subfigure}{.5\textwidth}
  \centering
  \includegraphics[width=0.9\linewidth]{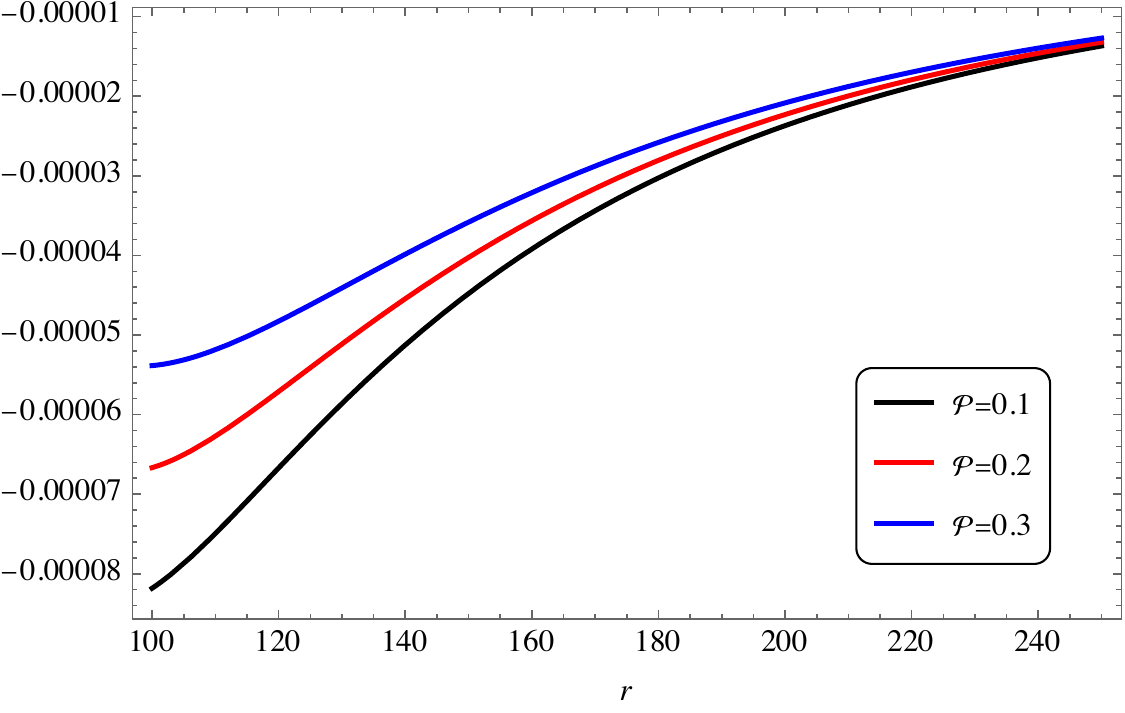}
\end{subfigure}
\caption{Left Panel: Energy density and its combinations with the pressures, as functions of the radial coordinate, for $r_0=100$, $a_0=1.0$, and $\mathcal{P}=0.1$. Right panel: Energy density plus radial pressure, in detail, for the same parameters and other values of $\mathcal{P}$.}
\label{Fig2}
\end{figure}

\begin{figure}[!h]
 \centering
   \includegraphics[width=0.6\textwidth]{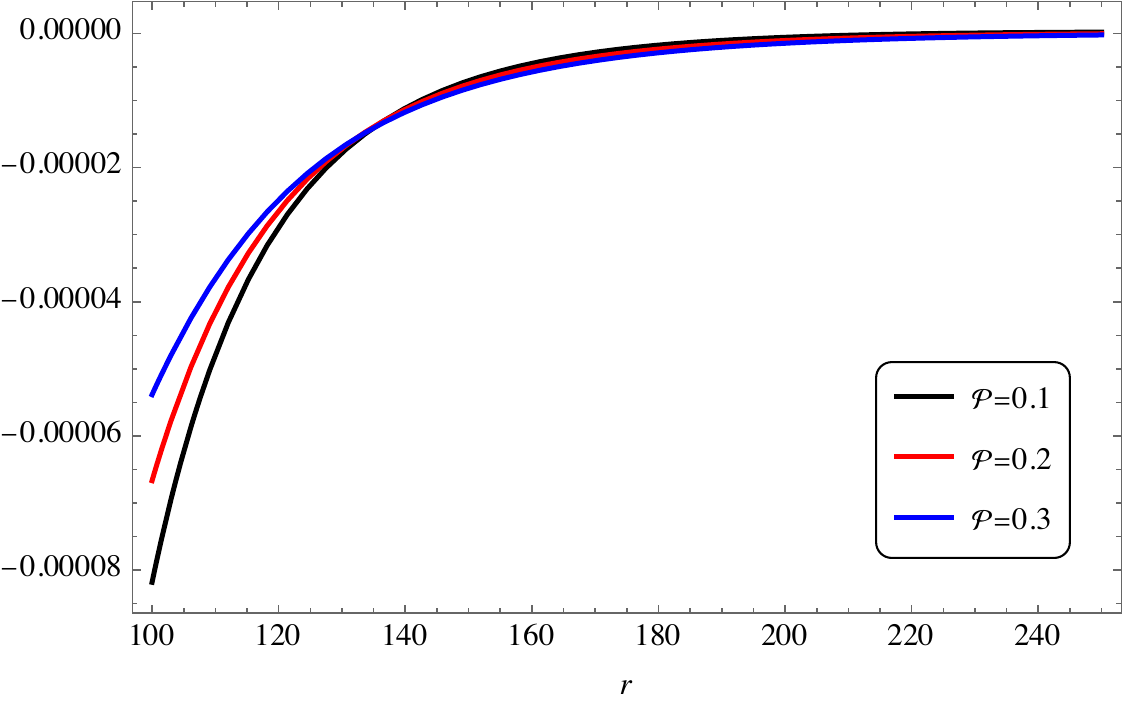}
      \caption{Energy density plus radial pressure, $\rho+p_r+2p_t$, for the same parameters $a_0=1.0, r_0=100$ and other values of $\mathcal{P}$.}
  \label{EnCond2}
\end{figure}

\begin{figure}[!h]
 \centering
u   \includegraphics[width=0.6\textwidth]{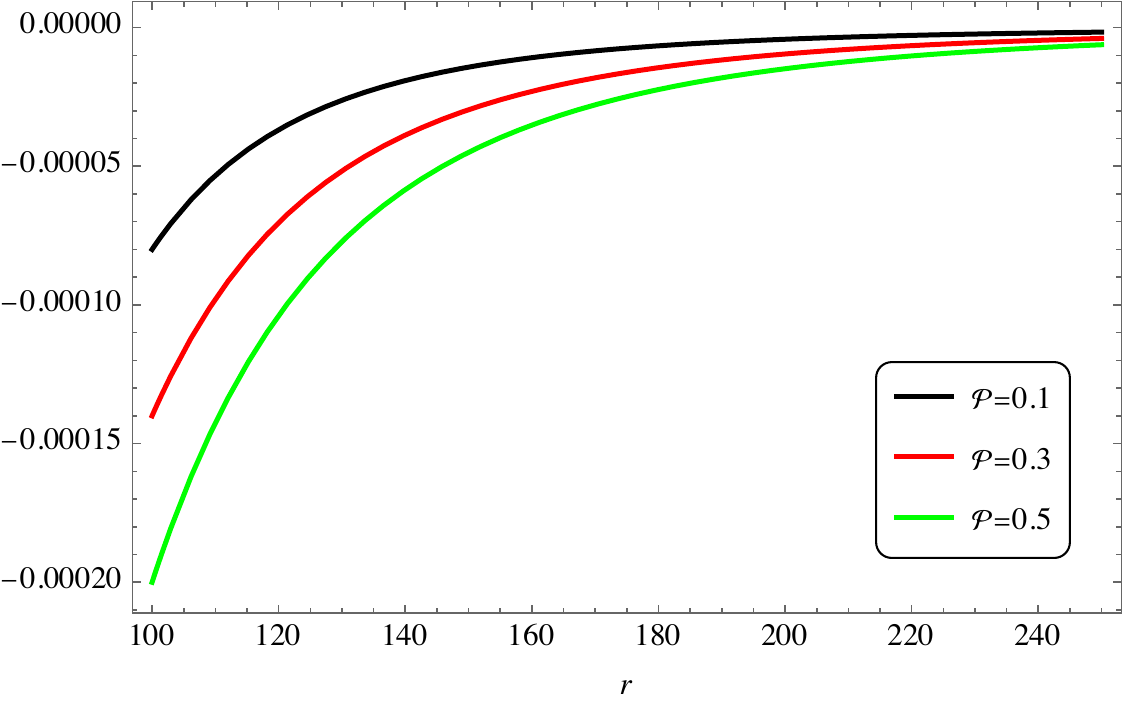}
      \caption{Ricci's curvature scalar as a function of the radial coordinate, for some values of $\mathcal{P}$, with $a_0=1.0$, $r_0=100$.}
  \label{RD}
\end{figure}

Regarding the VIQ (Volume Integral Quantifier) for this solution, it can be demonstrated that the amount of exotic matter necessary to maintain an open wormhole throat, especially for larger wormholes, aligns consistently with the previously examined scenarios. To be more specific, it converges to $-8 \pi \sqrt{a_0}(1-2\mathcal{P})$.

\begin{figure}[!h]
\centering
\begin{subfigure}{.5\textwidth}
  \centering
  \includegraphics[width=1.00\linewidth]{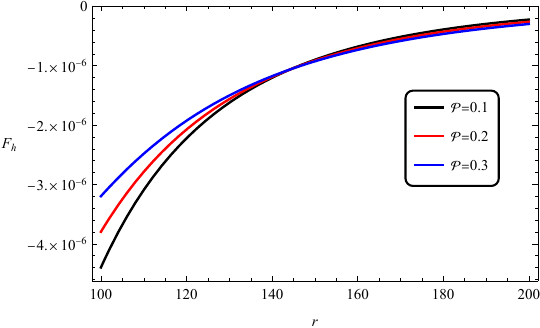}
\end{subfigure}%
\begin{subfigure}{.5\textwidth}
  \centering
  \includegraphics[width=1.00\linewidth]{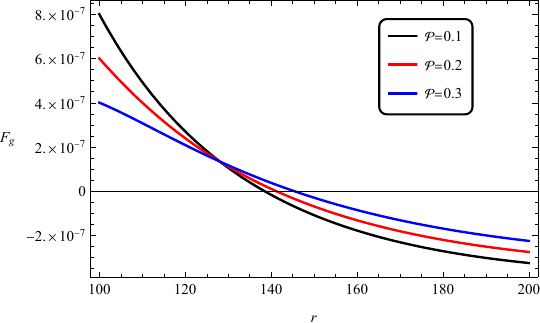}
\end{subfigure}
\begin{subfigure}{.5\textwidth}
  \centering
  \includegraphics[width=1.01\linewidth]{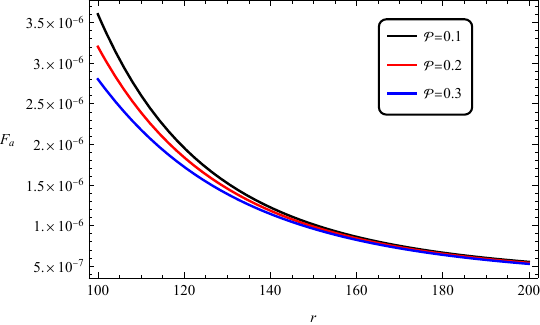}
\end{subfigure}
\caption{Left Panel (top): Hydrostatic force as a function of the radial coordinate, for $r_0=100$, $a_0=1.0$, and some values of $\mathcal{P}$. Right panel (top): Gravitational force, for the same values of the parameters. Bottom panel: Anisotropy force.}
\label{Forces}
\end{figure}

Finally, upon satisfying the TOV equation, we confirm the fluid's equilibrium state. In Figure \ref{Forces}, the left panel (top) illustrates the hydrostatic force $F_h=-dp_r/dr$ for various LQG parameter ($\mathcal{P}$) values. Meanwhile, the right panel (top) portrays the gravitational force, $F_g=-\Phi'(r)(\rho+p_r)$, for the same parameter values, both plotted against the radial coordinate. The bottom panel represents the anisotropy force, $F_a=(2/r)(p_t-p_r)$. Notably, these forces, particularly their magnitudes near the wormhole throat, decrease as the LQG parameter diminishes. As we have seen, the amplification of quantum fluctuations impacts the fabric of spacetime near the wormhole, affecting the matter that sustains it. Consequently, there is a reduction in the forces required to maintain the wormhole's integrity and stability. 

\subsection{Case $\Phi(r)=\frac{1}{2}\log{\left(1+\frac{\mathcal{P} \sqrt{a_0}}{r}\right)}$}

For the wormhole solution whose redshift function now depends explicitly on the LQG parameters, the obtained novel radial and transversal pressures are given by
\begin{eqnarray}\label{pr2}
p_r&=&-\frac{1}{{10 r^7 r_0^3 \left(\sqrt{a_0} \mathcal{P}+r\right)}}(2 a_0^2 \mathcal{P} r^5-70 a_0^2 \mathcal{P} r r_0^4+68 a_0^2 \mathcal{P} r_0^5-5 a_0^2 r^5\nonumber\\
&+&40 a_0^2 r^2 r_0^3-35 a_0^2 r r_0^4+10 \sqrt{a_0} \mathcal{P} r^5 r_0^3+20 \mathcal{P} r^5 r_0^4-20 \mathcal{P} r^4 r_0^5+10 r^5 r_0^4),\\ 
p_t&=&\frac{1}{40 r^8 r_0^3 \left(\sqrt{a_0} \mathcal{P}+r\right)^2}\left(\sqrt{a_0} \mathcal{P}+2 r\right) (-280 a_0^{5/2} \mathcal{P}^2 r r_0^4+340 a_0^{5/2} \mathcal{P}^2 r_0^5
+120 a_0^{5/2} \mathcal{P} r^2 r_0^3\nonumber\\
&-&140 a_0^{5/2}\mathcal{P} r r_0^4+2 a_0^2 \mathcal{P} r^6-350 a_0^2 \mathcal{P} r^2 r_0^4+408 a_0^2 \mathcal{P} r r_0^5-5 a_0^2 r^6+160 a_0^2 r^3 r_0^3-175 a_0^2 r^2 r_0^4\nonumber\\
&-&20 \sqrt{a_0} \mathcal{P}^2 r^4 r_0^5+10 \sqrt{a_0} \mathcal{P} r^6 r_0^3+20 \mathcal{P} r^6 r_0^4-40 \mathcal{P} r^5 r_0^5+10 r^6 r_0^4)\label{pt2}.
\end{eqnarray}
\begin{figure}[!h]
 \centering
   \includegraphics[width=0.6\textwidth]{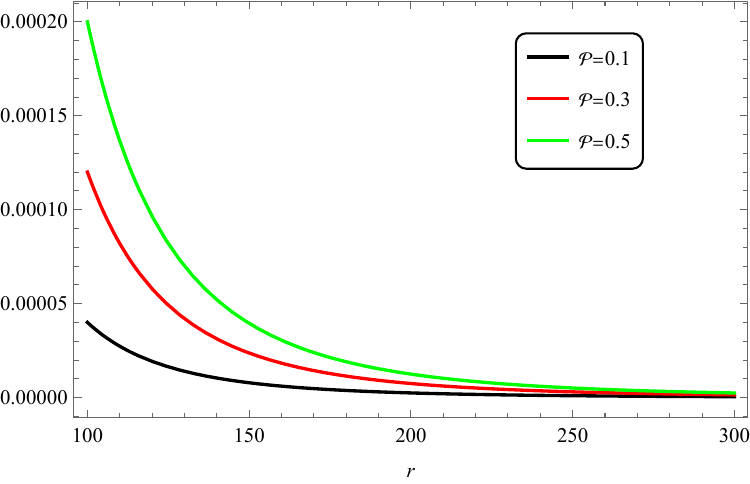}
      \caption{Ricci's curvature scalar as a function of the radial coordinate, for some values of $\mathcal{P}$, with $a_0=1$ and $r_0=100$.}
  \label{R2}
\end{figure}
\begin{figure}[!h]
 \centering
   \includegraphics[width=0.6\textwidth]{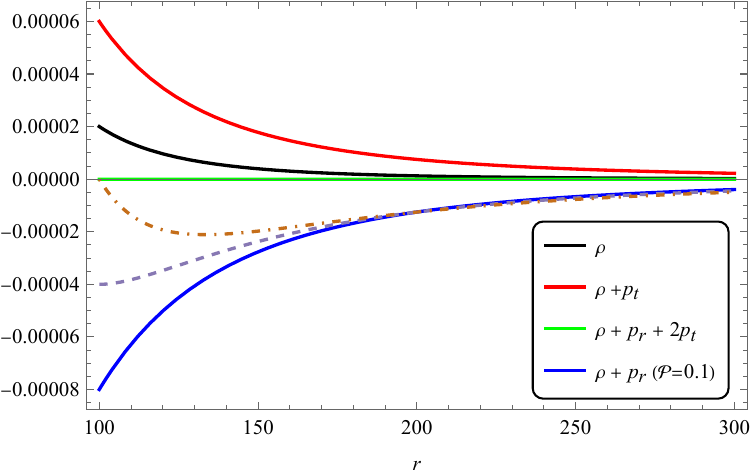}
      \caption{Full lines: Energy density and its combinations with the pressures, as functions of the radial coordinate, for $r_0=100$, $a_0=1.0$, and $\mathcal{P}=0.1$. Dashed and dot-dashed lines: $\rho+p_r$, for $\mathcal{P}=0.3$ and $\mathcal{P}=0.5$, respectively, considering the same parameters $a_0$ and $r_0$.}
  \label{EnCond2}
\end{figure}

The first feature of this wormhole solution that we will analyze is its Ricci curvature scalar. In Figure \ref{R2}, we depict this quantity as a function of the radial coordinate for some values of $\mathcal{P}$. Note that this wormhole solution exhibits increased curvature with higher LQG parameters. As we have seen, this parameter influences spacetime by introducing pronounced quantum effects. As they intensify, the interplay between classical spacetime and emerging quantum characteristics becomes more significant, leading to a stronger deviation from classical geometry and a more accentuated curvature.

Examining the energy conditions according to Figure \ref{EnCond2}, which was built from Eqs. (\ref{rho}), (\ref{pr2}), and (\ref{pt2}), it is evident that all of them are violated due to the non-fulfillment of the Null Energy Condition (NEC). This violation diminishes as the $\mathcal{P}$ parameter increases, indicating an attenuation in the violation trend. Remarkably, at the critical value $\mathcal{P}=1/2$, all conditions are met at the wormhole throat, a distinct contrast from the behavior observed in the zero-tidal solution.

Concerning the VIQ for this solution, it's demonstrated that the quantity of exotic matter required to sustain an open wormhole throat, particularly for large wormholes, remains consistent with previously analyzed cases. Specifically, it converges to $\mathcal{I}_q\to -8\pi\sqrt{a_0}(1-2\mathcal{P})$.

\begin{figure}[!h]
\centering
\begin{subfigure}{.5\textwidth}
  \centering
  \includegraphics[width=1.01\linewidth]{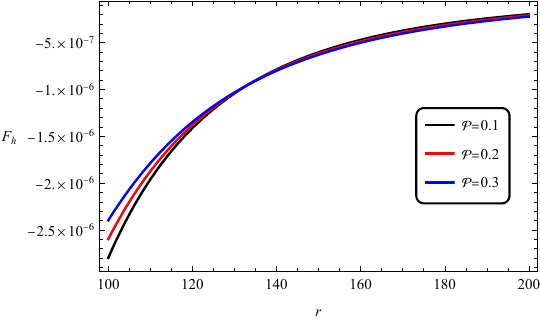}
\end{subfigure}%
\begin{subfigure}{.5\textwidth}
  \centering
  \includegraphics[width=1.01\linewidth]{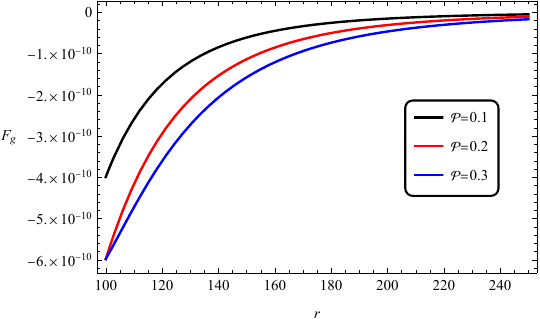}
\end{subfigure}
\begin{subfigure}{.5\textwidth}
  \centering
  \includegraphics[width=1.01\linewidth]{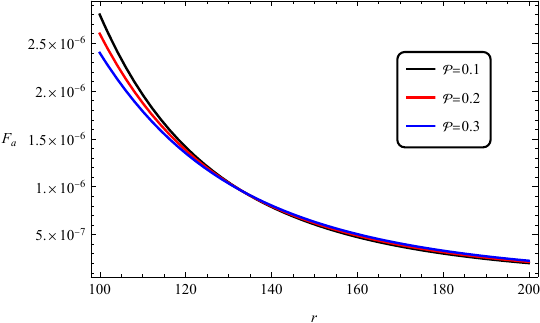}
\end{subfigure}
\caption{Left Panel (top): Hydrostatic force as a function of the radial coordinate, for $r_0=100$, $a_0=1.0$, and some values of $\mathcal{P}$. Right panel (top): Gravitational force, for the same values of the parameters. Bottom panel: Anisotropy force.}
\label{Forces1}
\end{figure}

The equilibrium of this wormhole solution is examined via the TOV equation by analyzing hydrostatic, gravitational, and anisotropy forces, which are plotted against the radial coordinate in Figure \ref{Forces1}. The top left panel illustrates the hydrostatic force for varying $\mathcal{P}$ values, while the top right panel shows the gravitational force; the bottom panel represents the anisotropy force. Once more, near the wormhole throat, the magnitudes of hydrostatic ($F_h$) and anisotropy ($F_a$) forces diminish as $\mathcal{P}$ increases, consistent with previous findings. Conversely, the absolute value of the gravitational force increases with higher LQG parameter values. This gravitational force depends on the redshift function's derivative, which also is influenced by the LQG parameters in this wormhole solution. Greater LQG parameters correspond to heightened quantum fluctuations of the spacetime, in this case intensifying this force to uphold the structure and stability of the wormhole.

\subsection{Case of the fluid with EoS $p_r=\omega\rho$}

We will analyze now the wormhole solution considering that the source obeys this equation of state. Einstein`s equation (\ref{eq:grr}) cannot be analytically solved for $\Phi(r)$ given Eqs. (\ref{SolApp}) and (\ref{rho}), then we expand its coefficients for small LQG parameters, as we have considered so far, and thus we manage to integrate it. Terms of $\mathcal{O}(1/r)$ and upper will arise, besides the term
$30 \left[2 a_0^2 (\mathcal{P}+1) (\omega+1)+r_0^4 (2 \mathcal{P} (\omega+1)+1)\right]\log (1-r_0/r) $, indicating that the solution will diverge for $r=r_0$. Therefore, we will impose that the coefficient of this term must vanish to prevent the appearance of horizons, and thus we arrive at the following expression for the state parameter:
\begin{equation}\label{omega}
    \omega=-\frac{2a_0^2(\mathcal{P}+1)+r_0^4(2\mathcal{P}+1) }{2 \left[a_0^2 (\mathcal{P}+1)+\mathcal{P} r_0^4\right]},
\end{equation}
where $\omega\approx-1 -r_0^4/2a_0^2$, for $\mathcal{P}\to 0$. Therefore, considering the macroscopic wormholes that we are investigating here, the fluid is a phantom-type, since $\omega<-1$. 

Taking into account the elimination of the logarithmic term from the constraint for $\omega$ given by Eq. (\ref{omega}), the redshift function becomes
\begin{eqnarray}
    \Phi(r)&=&\frac{a_0^2 \mathcal{P} r_0 \omega}{10 r^5}+\frac{a_0^2}{r^4}\left(\frac{\mathcal{P} \omega}{2}  +\frac{9  \mathcal{P}}{10}+\frac{7 w}{4}\right)+\frac{a_0^2}{r_0 r^3}\left(\frac{ \mathcal{P} \omega}{2}+\frac{9 \mathcal{P}}{10}+\frac{\omega}{3}+\frac{7}{12}\right)\nonumber\\
    &+&\frac{a_0^2}{r_0^2r^2}\left(\frac{ \mathcal{P} \omega}{2}+\frac{9\mathcal{P}}{10}+\frac{ \omega}{2}+\frac{3}{4}\right)+\frac{a_0^2}{r_0^3 r}\left(\mathcal{P} \omega+\frac{\mathcal{P} r_0^4\omega}{a_0^2}+\frac{9 \mathcal{P}}{10}+\omega+\frac{5}{4}\right).
\end{eqnarray}
 It can be shown that as the LQG parameters increase, the redshift function, impacting the passage of light and signals through the wormhole, also rises, affecting thus the gravitational conditions experienced when traversing it. We also remark that when those parameters vanish the redshift function is the same analyzed in the subsection B.

In Figure \ref{RD}, the graph illustrates the Ricci curvature scalar as a function of $r$. It is remarkable that the parameter $\mathcal{P}$ directly influences the trend of increasingly spacetime curvature. This consistent pattern emphasizes the pivotal role played by spacetime's quantum properties in accentuating its curvature, as observed in prior analysis. It is worth noting that in the case under consideration, for lower values of that LQG parameter, the curvature becomes negative. 
\begin{figure}[!h]
 \centering
   \includegraphics[width=0.6\textwidth]{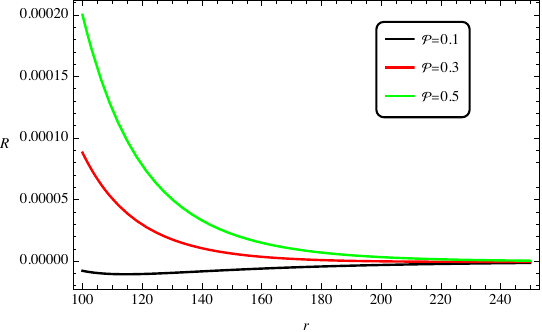}
      \caption{Ricci's curvature scalar as a function of the radial coordinate, for some values of $\mathcal{P}$, with $a_0=1$, $r_0=100$, and $\omega$ obeying Eq. (\ref{omega}).}
  \label{RD}
\end{figure}

Due to the source being a phantom-type fluid, it results in the violation of all energy conditions. Specifically, the null energy condition, expressed as $\rho(1+\omega)\geq 0$, fails to hold. Observing the expression for $\omega$ in equation (\ref{omega}), it's evident that for fixed values of $a_0$ and $r_0$, the modulus of state parameter decreases as $\mathcal{P}$ increases. Consequently, this escalation of $\mathcal{P}$ once again diminishes the severity of the energy conditions violation.

Concerning VIQ analysis for large wormholes, the obtained value near the throat is given by $\mathcal{I}_q=-8 \pi\sqrt{a_0}$ on considering Eq. (\ref{omega}). This finding stands independent of alterations in $\mathcal{P}$, distinguishing it from previous trends. Importantly, this denotes that the necessary amount of exotic matter at the wormhole's throat remains unaffected by the underlying quantum properties of spacetime.

\section{Conclusions}

This study investigated traversable Lorentzian wormhole solutions derived from Loop Quantum Gravity (LQG) theory. It revealed novel static and spherically symmetric Morris-Thorne solutions that share an energy source similar to the one of the self-dual regular black hole discussed in \cite{Modesto:2009ve}. The main focus was on examining how these solutions and their behavior depend on two critical LQG parameters, namely $a_0$ and $\mathcal{P}$, which govern the quantum properties of spacetime. Specifically, the analysis concentrated on exploring the sustainability of macroscopic wormholes under the influence of small values of these parameters.

The exploration initiated with the simplest scenario, the zero-tidal traversable wormhole, with a constant redshift function (case A). A solution was derived from Einstein's equations, revealing that the shape function -- furthermore, valid for all the wormhole solutions studied here -- satisfies the conditions essential for its viability, including asymptotic flatness and the flaring-out condition, provided the LQG parameters rely on a range of values, which aligns conveniently with the utilized approximation. The embedding diagrams were plotted, indicating an increasingly pronounced curvature towards the wormhole throat as the LQG parameter $\mathcal{P}$ is augmented. On the other hand, analysis of the Ricci scalar revealed decreasing curvatures with larger values of that parameter for the case under consideration.

Regarding still case A, the energy conditions (basically, WEC and SEC) were graphically explored, revealing that the increasing of the LQG parameter $\mathcal{P}$, linked to spacetime's quantum behavior, notably reduces the obtained violations of these conditions. This observation suggests that quantum effects, represented by the LQG parameter, might mitigate violations in the energy conditions essential for the wormhole's feasibility.  Consistently, the investigations of the Volume Integral Quantifier (VIQ) in the immediate throat vicinity have indicated a reduction in the requirement for exotic matter as $\mathcal{P}$ intensifies. We also verified that the TOV equation, which represents the equilibrium of the material source, is consistently satisfied.

The subsequent analysis examined some wormhole solutions with non-constant redshift function: one independent of LQG parameters ($\Phi=r_0/r$, case B), another explicitly depending on these parameters ($\Phi=1/2 \log{(1+\mathcal{P} \sqrt{a_0}/r)}$, case C), and one adhering to the EoS $p_r=\omega \rho$ (case D). Concerning this latter, the redshift function depends on the LQG parameters as well as on the state parameter, $\omega$. Besides this, the condition for the absence of horizons imposed a constraint in this latter, so that to have a macroscopic wormhole the fluid must be a phantom-type. The investigation demonstrated that the solution with the LQG parameter-independent redshift function, case B, presents behavior similar to the zero-tidal solution, reducing Ricci's curvature as $\mathcal{P}$ increases. Conversely, the other solutions exhibit an inverse trend, with increased spacetime quantum effects magnifying the curvature. 

The violation of the energy conditions for all the solutions with non-constant redshift also became attenuated with the increasing of $\mathcal{P}$. Regarding VIQ, only the solution satisfying EoS presented an amount of exotic matter near the wormhole throat independent of this LQG parameter. The other solutions share the same behavior as the zero-tidal one, reducing the requirement of exotic matter near the throat with the increasing of $\mathcal{P}$. 

Finally, examining equilibrium conditions using the TOV equation for cases B and C unveiled distinctions between them. In case B, the gravitational force magnitude increases with $\mathcal{P}$, while the other forces decrease. In contrast, for case C, all forces diminish as the parameter increases, indicating a general softening in the fluid's equilibrium conditions with the intensification of quantum effects in spacetime.

Each scenario here studied presented nuanced insights into how LQG-inspired wormholes respond to diverse conditions, shedding light on their stability, curvature, and adherence to critical physical conditions. The interplay between classical spacetime and emerging quantum effects appears to significantly influence these wormholes' structural integrity and behavior.

\section*{Acknowledgments}
\hspace{0.5cm} CRM thanks the Conselho Nacional de Desenvolvimento Cient\'{i}fico e Tecnol\'{o}gico (CNPq), Grants no. 308268/2021-6.




\begin{thebibliography}{99}

\bibitem{Oppenheimer:1939ue}
J.~R.~Oppenheimer and H.~Snyder,
Phys. Rev. \textbf{56}, 455-459 (1939).

\bibitem{Hawking:1988ae}
S.~W.~Hawking,
Phys. Rev. D \textbf{37}, 904-910 (1988).

\bibitem{Morris:1988cz}
M.~S.~Morris and K.~S.~Thorne,
Am. J. Phys. \textbf{56}, 395-412 (1988).

\bibitem{Einstein:1935tc}
A.~Einstein and N.~Rosen,
Phys. Rev. \textbf{48}, 73-77 (1935).

\bibitem{Morris:1988tu}
M.~S.~Morris, K.~S.~Thorne and U.~Yurtsever,
Phys. Rev. Lett. \textbf{61}, 1446-1449 (1988).

\bibitem{Frolov:2023res}
V.~P.~Frolov, P.~Krtous and A.~Zelnikov,
Phys. Rev. D \textbf{108}, no.2, 024034 (2023).

\bibitem{LIGOScientific:2016aoc}
B.~P.~Abbott \textit{et al.} [LIGO Scientific and Virgo],
Phys. Rev. Lett. \textbf{116}, no.6, 061102 (2016).

\bibitem{Rovelli:2014ssa}
C.~Rovelli and F.~Vidotto,
\textit{Covariant Loop Quantum Gravity: An Elementary Introduction to Quantum Gravity and Spinfoam Theory},
Cambridge University Press, 2014.

\bibitem{Rovelli:2003wd}
C.~Rovelli,
Int. J. Mod. Phys. D \textbf{12}, 1509-1528 (2003).

\bibitem{Zwiebach:2004tj}
B.~Zwiebach,
\textit{A first course in string theory}, Cambridge University Press, 2004.

\bibitem{Modesto:2005zm}
L.~Modesto,
Class. Quant. Grav. \textbf{23}, 5587-5602 (2006).

\bibitem{Peltola:2008pa}
A.~Peltola G.~Kunstatter,
Phys. Rev. {\bf D79}, 061501 (2009).

\bibitem{Gambini:2013ooa}
R.~Gambini and J.~Pullin,
Phys. Rev. Lett. \textbf{110}, no.21, 211301 (2013).

\bibitem{Olmedo:2016ddn}
J.~Olmedo,
Universe \textbf{2}, no.2, 12 (2016).

\bibitem{Sengupta:2023yof}
R.~Sengupta, S.~Ghosh and M.~Kalam,
Eur. Phys. J. C \textbf{83}, no.9, 830 (2023).
   
\bibitem{Modesto:2009ve}
L.~Modesto and I.~Premont-Schwarz,
Phys. Rev. D \textbf{80}, 064041 (2009).

\bibitem{Brown:2010csa}
E.~Brown, R.~B.~Mann and L.~Modesto,
Phys. Lett. B \textbf{695}, 376-383 (2011).

\bibitem{Silva:2017gki}
C.~A.~S.~Silva and F.~A.~Brito,
Universe \textbf{3}, no.2, 42 (2017).

\bibitem{Cruz:2015bcj}
M.~B.~Cruz, C.~A.~S.~Silva and F.~A.~Brito,
Eur. Phys. J. C \textbf{79}, no.2, 157 (2019).

\bibitem{Cruz:2020emz}
M.~B.~Cruz, F.~A.~Brito and C.~A.~S.~Silva,
Phys. Rev. D \textbf{102}, no.4, 044063 (2020).

\bibitem{Santos:2021wsw}
J.~S.~Santos, M.~B.~Cruz and F.~A.~Brito,
Eur. Phys. J. C \textbf{81}, no.12, 1082 (2021).

\bibitem{Fuller:1962zza}
R.~W.~Fuller and J.~A.~Wheeler,
Phys. Rev. \textbf{128}, 919-929 (1962).

\bibitem{SupernovaSearchTeam:1998fmf}
A.~G.~Riess \textit{et al.} [Supernova Search Team],
Astron. J. \textbf{116}, 1009-1038 (1998).

\bibitem{Dalal:2000xw}
N.~Dalal and K.~Griest,
Phys. Lett. B \textbf{490}, 1-5 (2000).

\bibitem{Cai:2009zp}
Y.~F.~Cai, E.~N.~Saridakis, M.~R.~Setare and J.~Q.~Xia,
Phys. Rept. \textbf{493}, 1-60 (2010).

\bibitem{Yang:2023gas}
S.~Yang, W.~Guo, Q.~Tan, and Y.~Liu,
 Phys. Rev. {\bf D108}, 2, 024055 (2023).

\bibitem{Zlatev:1998tr}
I.~Zlatev, L.~M.~Wang and P.~J.~Steinhardt,
Phys. Rev. Lett. \textbf{82}, 896-899 (1999).

\bibitem{Urena-Lopez:2002nup}
L.~A.~Urena-Lopez and A.~R.~Liddle,
Phys. Rev. D \textbf{66}, 083005 (2002).

\bibitem{Dvali:2000hr}
G.~R.~Dvali, G.~Gabadadze and M.~Porrati,
Phys. Lett. B \textbf{485}, 208-214 (2000).

\bibitem{Nojiri:2006ri}
S.~Nojiri and S.~D.~Odintsov,
eConf \textbf{C0602061}, 06 (2006).

\bibitem{Das:2022wzp}
P.~Das and M.~Kalam,
Eur. Phys. J. C \textbf{82}, no.4, 342 (2022).

\bibitem{Estrada:2023pny}
M.~Estrada and C.~R.~Muniz, 
JCAP {\bf03}, 055 (2023).

\bibitem{Santos:2023zrj}
A.~L.~Santos, C.~R.~Muniz, and R.~V.~Maluf,
 JCAP {\bf09}, 022 (2023).
 
\bibitem{Barcelo:1999hq}
C.~Barcelo and M.~Visser,
Phys. Lett. B \textbf{466}, 127-134 (1999).

\bibitem{Hayward:2002pm}
S.~A.~Hayward,
Phys. Rev. D \textbf{65}, 124016 (2002).

\bibitem{Sengupta:2023ysx}
R.~Sengupta, S.~Ghosh, B.~C.~Paul and M.~Kalam,
Class. Quant. Grav. \textbf{40}, no.9, 095009 (2023).

\bibitem{Eiroa:2005pc}
E.~F.~Eiroa and C.~Simeone,
Phys. Rev. D \textbf{71}, 127501 (2005).

\bibitem{Rosa:2022osy}
J.~L.~Rosa and P.~M.~Kull,
Eur. Phys. J. C \textbf{82}, no.12, 1154 (2022).

\bibitem{Muniz:2022eex}
C.~R.~Muniz and R.~V.~Maluf, 
Annals Phys. {\bf446}, 169129 (2022).

\bibitem{Ashtekar:2008ay}
A.~Ashtekar,
J. Phys. Conf. Ser. \textbf{189}, 012003 (2009).

\bibitem{Nandi:2004}
K. K. Nandi, Y. Z. Zhang and K. B. Kumar,
Phys. Rev. D \textbf{70}, 127503 (2004).

\bibitem{Bonanno:2000ep}
A.~Bonanno and M.~Reuter,
Phys. Rev. D \textbf{62}, 043008 (2000).

\end{thebibliography}
\end{document}